\documentstyle[titlepage,12pt,a4]{article}

\begin{document}
\thispagestyle{empty}
\begin{center}
{\Large\bf Significance of zero modes in path--integral quantization of
solitonic theories with BRST invariance}
\\[14mm]
{\large J.--G. Zhou\raisebox{0.8ex}{\small a},
F. Zimmerschied\raisebox{0.8ex}{\small a,}\footnote{E--mail:
zimmers@physik.uni--kl.de},
J.--Q. Liang\raisebox{0.8ex}{\small a,b},\\
H.~J.~W. M\"uller--Kirsten\raisebox{0.8ex}{\small a} and
D.~H. Tchrakian\raisebox{0.8ex}{\small a,c}}
\\[1cm]
{\it \raisebox{0.8ex}{\small a} Department of Physics \\
University of Kaiserslautern, P.\ O.\ Box 3049, D 67653 Kaiserslautern,
Germany \\[5mm]
\raisebox{0.8ex}{\small b} Institute of Theoretical Physics, \\
Shanxi University Taiyuan, Shanxi 03006, P.\ R.\ China \\
and \\
Institute of Physics \\
Academia Sinica, Beijing 100080, P.\ R.\ China \\[5mm]
\raisebox{0.8ex}{\small c} Department of Mathematical Physics \\
St.\ Patrick's College, Maynooth, Ireland}
\\ \vfill
{\bf Abstract}
\end{center}
The significance of zero modes in the path--integral quantization of some
solitonic models is investigated. In particular a Skyrme--like theory
with topological vortices in $(1+2)$ dimensions is studied, and with a BRST
invariant gauge fixing a well defined transition amplitude is obtained in the
one loop approximation. We also present an alternative
method which does not necessitate evoking
the time--dependence in the functional integral, but
is equivalent to the original one in dealing with the quantization
in the background of the static classical solution of
the non--linear field equations. The considerations given here
are particularly useful in -- but also limited to --the
one--loop approximation.
\newpage
\section{Introduction}

The solutions of static (time--independent) non--linear classical
Euler--Lagrange equations are of particular interest in many field theories.
The energy of the static classical configuration (named soliton) is higher than
the minimum energy of a constant field (the perturbation theory vacuum) by a
finite amount. Interpreting the space coordinate $x$ as a Euclidean time in
$(1+1)$ dimensional field equations the static classical configuration is
the same as the $(1+0)$ dimensional
instanton which is responsible for quantum tunneling through barriers
\cite{1}--\cite{4}. From translation invariance in $(1+1)$ dimensions it is
clear that the solution of the static equation is a function of $x-a$ where
$a$ is an arbitrary constant of integration, corresponding to the origin
of some reference frame or the position of the soliton. We may set the
constant $a$ to zero since no physically interesting quantity can depend on it
\cite{5}. However, because of the translation invariance, functional integrals, i.\
e.\ transition amplitudes in the path--integral formulation are not well defined
as can be seen by recalling that the second variation of the action $S$ about a
static solution leads to zero modes which give rise to undefined integrals in the
perturbation expansion about the soliton. In order to cure this problem one
often resorts to the so--called Faddeev--Popov technique \cite{6} by inserting
an identity for $1$ with a $\delta$--function integral which transforms the integral
over a zero mode into a continous integral over the translation parameter $a$
or the position of the soliton.  In this procedure much care has to be
taken in calculating the Jacobian of the transformation.
In the collective coordinate method with BRST
invariance \cite{7} -- \cite{10} the shift of the integration variable related to
the zero mode is achieved in a natural way by regarding the position of the
soliton as a new dynamical variable, namely the collective coordinate, which
depends on time. The static solution thus becomes time--dependent through
the collective coordinate. Generally speaking, whenever the action of a static
field possesses some symmetry the operator of the second variation of the
action about the classical static solution has corresponding zero modes \cite{11,12}.
The BRST invariant gauge fixing breaks these symmetries and gives rise to
well defined functional integrals. However, applications of the method to
specific models are scarce since only very few permit explicit calculations.  Below
we consider two such models with the intent to expose in particular
the vital role played by zero modes.

In the following we  therefore investigate the path--integral quantization of 
some solitonic models and begin with a
theory with a Skyrme--like soliton with absolute scale in $(1+2)$ dimensions
\cite{12}. The model
--- motivated by Skyrme--term--modified $\sigma$ models \cite{13} and
useful e.\ g.\ as the skeleton of a superconducting cosmic string
\cite{14} ---
involves a complex scalar field and lends itself for some aspects more
readily as a less trivial testing ground of the quantization method than
the usual soliton model in $(1+1)$ dimensions.
In spite of its drawback of being nonrenormalizable in the usual sense
it has some intriguing positive aspects which make it a useful laboratory for the
study of various phenomena.  Thus the model allows the explicit solution of
the classical static equation and the explicit demonstration of the
topological stability of the solutions.
The nonzero topological charge of these static classical solutions
provides a lower bound
to the energy integral and renders the solutions stable. But it is also possible
to find nontopological static solutions with finite energy which are classically
unstable \cite{15}.
The topological vortex--type solutions of the model are very similar
to those of the Nielsen--Olesen model \cite{16}
but are generated by the same complex scalar field and can be 
calculated explicitly in closed form \cite{12}, which makes the model
in any case an interesting laboratory.  However, the model has another
fascinating property.  Being a model in $(2+1)$ dimensions it can be
supersymmetrized by a well--known method \cite{17}. $(2+1)$ dimensions
is the lowest dimensionality for which the topological charge appears as the central
charge of an $N=2$ extended supersymmetry of the $N=1$ supersymmetrized
theory.  The requirement that a theory with a topological charge
have a consistent superymmetric extension implies that the theory 
exhibits a Bogomol`nyi relation \cite{18} which can be shown to be the case.  Our 
Skyrme--like model is a model with these properties.
In the case of this Skyrmion theory \cite{12} to be considered below the
action has not only the usual translation symmetry but also a global $U(1)$
symmetry which together lead to three zero modes of the second variation
operator about the classical configuration and hence to an interesting
set of constraints. Though the kinetic energy term is
quartic the constraints  induced by introducing collective coordinates are seen
to be primary and first class. With the help of a BRST invariant gauge fixing a
well defined transition amplitude is given up to the one loop approximation.

We then show  in the context of a different model
that it is not necessary to evoke the time dependence in the
functional integrals by regarding transformation parameters as
time--dependent dynamical variables. Expanding the action about the classical
static--field configuration the classical action can be factored out. The
remaining fluctuation part of the action possesses a new and
very interesting symmetry under
a shift by zero modes (analogous to observations made in
\cite{19}). This symmetry can also be broken with BRST invariant
gauge fixing to give the desired transition amplitude. The two treatments are seen
to be equivalent in the one loop approximation, and the latter is restricted to this.

We add that the significance of zero modes has been discussed in particular
by Peskin \cite{20} and Bernard \cite{21}. The BRST quantization of a soliton
in $(2+1)$ dimensions has recently been discussed in \cite{22}. General considerations of
collective coordinates are discussed in \cite{23}.

\section{Symmetry of the Skyrme--like soliton theory, collective coordinates
and zero modes}

The model we consider first is defined by the following Lagrangian density
in $(1+2)$ dimensional Minkowski space \cite{12} ($\eta_{00}=1, \eta_{ij}=
-\delta_{ij}$),
\begin{eqnarray}
{\cal L} & = & \frac{1}{2} (i\partial_{[\mu}\varphi\partial_{\nu]}\varphi^\ast)
(i\partial^{[\nu}\varphi\partial^{\mu]}\varphi^\ast) - U(\varphi
\varphi^\ast) \nonumber \\
& = & - (\partial_0\varphi\partial_i\varphi^\ast-\partial_i\varphi\partial_0
\varphi^\ast)^2+ \frac{1}{2}
(\partial_i\varphi\partial_j\varphi^\ast-\partial_j\varphi\partial_i
\varphi^\ast)^2-U(\varphi\varphi^\ast)
\label{2.1}
\end{eqnarray}
The finite--energy soliton--like solutions we are interested in are the static
field configurations $\varphi_c,\varphi_c^\ast$ (with $\partial_0\varphi_c=
\partial_0\varphi_c^\ast=0$) which minimize the energy, namely the static Hamiltonian
\begin{equation}
H_0=\int{\cal H}_0 d^2x = - \int{\cal L}_0 d^2x = S_0
\label{2.2}
\end{equation}
where the Lagrangian density of the static version is
\begin{equation}
{\cal L}_0=\frac{1}{2}
(\partial_i\varphi\partial_j\varphi^\ast-\partial_j\varphi\partial_i\varphi^\ast)^2
-U(\varphi\varphi^\ast)
\label{2.3}
\end{equation}
and
\begin{equation}
U(\varphi\varphi^\ast)=\lambda^2(\eta^2-\varphi\varphi^\ast)^{2k},
\end{equation}
$k>1$ an integer.
The finite energy soliton--like fields $\varphi_c,\varphi_c^\ast$  with vortex shape studied in
reference \cite{12} are
\begin{equation}
\varphi_c=\eta R(r) e^{in\theta}, \qquad
\varphi_c^\ast =\eta R(r) e^{-in\theta}
\label{2.4}
\end{equation}
and represent maps from $S^1$ to $S^1$, thus belonging to equivalence
classes of the fundamental homotopy group $\Pi_1(S^1)$ which are
characterized by an integer $n$, the homotopy charge.

The action $S_0 = \int {\cal L}_0 d^2x$ is invariant under translations of the
origin of the coordinate system, $\vec{x}\rightarrow\vec{x}+\vec{a}$, and is
also invariant under a global $U(1)$ transformation $\varphi\rightarrow
e^{i\alpha}\varphi,\varphi^\ast\rightarrow e^{-i\alpha}\varphi^\ast$. In view
of the associated degeneracy of the action, functional integrals representing
e.\ g.\ the transition amplitude, are not well defined since the symmetries
result in zero modes of the second variation operator of the action about
$\varphi_c$. These zero modes can be analysed in a general context. The
classical solutions $\varphi_c, \varphi_c^\ast$, of course, are obtained by
minimizing the action $S_0$,
\begin{equation}
\left.\frac{\delta S_0}{\delta\varphi}\right|_{\varphi=\varphi_c}\delta\varphi +
\left.\frac{\delta S_0}{\delta\varphi^\ast}
\right|_{\varphi^\ast=\varphi_c^\ast}\delta\varphi^\ast=0
\label{2.5}
\end{equation}
which leads to the equations of motion with solutions $\varphi_c,
\varphi_c^\ast$.

If the action $S_0$ as well as the equation of motion (\ref{2.5}), possess some
symmetries, namely, if they are invariant under correspondig transformations
$\varphi\rightarrow\varphi'=\varphi+\delta\varphi$, the second variation for the
zero modes $\delta\Phi_0=(\delta\varphi_0,\delta\varphi_0^\ast)$ vanishes, i.\ e.\
\begin{equation}
\int d^2xd^2y \delta\Phi_0 \hat{M}(\varphi_c, \varphi_c^\ast)
\delta^2(x-y)\delta\Phi_0(\eta) = 0
\label{2.6}
\end{equation}
where
\begin{equation}
\hat{M}(\varphi_c, \varphi_c^\ast)=
\left(
\begin{array}{cc}
\left[\left.\frac{\delta^2 S_0}{\delta\varphi\delta\varphi}
\right|_{\varphi=\varphi_c}\right] &
\left[\left.\frac{\delta^2 S_0}{\delta\varphi\delta\varphi^\ast}
\right|_{\varphi = \varphi_c \atop
\varphi^\ast = \varphi_c^\ast}
\right] \\
\left[\left.\frac{\delta^2 S_0}{\delta\varphi^\ast\delta\varphi}
\right|_{\varphi = \varphi_c \atop
\varphi^\ast  = \varphi_c^\ast}
\right] &
\left[\left.\frac{\delta^2 S_0}{\delta\varphi^\ast\delta\varphi^\ast}
\right|_{\varphi^\ast=\varphi_c^\ast}\right]
\end{array}
\right).
\label{2.7}
\end{equation}
The elements of the operator $\hat{M}$ are given by
\begin{eqnarray}
\left[\left.\frac{\delta^2 S_0}{\delta\varphi\delta\varphi}
\right|_{\varphi=\varphi_c}\right] & = &
- 2 \left[\partial_i(\delta_{il}^{jk}\partial_k
\varphi_c^\ast\partial_l\varphi_c^\ast\partial_j)
-\lambda^2 k (2k-1)(\eta^2-|\varphi_c|^2)^{2(k-1)}
\varphi_c^\ast\varphi_c^\ast\right] \nonumber \\
\label{2.8} \\
\left[\left.\frac{\delta^2 S_0}{\delta\varphi\delta\varphi^\ast}
\right|_{\varphi = \varphi_c \atop
\varphi^\ast = \varphi_c^\ast}
\right] & = &
-2 \left\{\partial_i\left[(\delta_{il}^{kj}+\delta_{ij}^{kl})
\partial_k\varphi_c\partial_l\varphi_c^\ast\partial_j\right] \right.
\nonumber \\
&  & +{}\left.\lambda^2 k (\eta^2-|\varphi_c|^2)^{2(k-1)}
(\eta^2-2k|\varphi_c|^2)\right\}
\label{2.9}
\end{eqnarray}
where $\delta_{il}^{jk}=\delta_i^j\delta_l^k-\delta_i^k
\delta_l^j$. The other two elements of (\ref{2.7}) are
obtained by complex conjugation of (\ref{2.8}) and
(\ref{2.9}). Therefore the operator $\hat{M}$ of the second
variation has zero modes
\begin{equation}
\Psi_0^i = \left(\frac{\partial\varphi_c}{\partial a_i},
\frac{\partial\varphi_c^\ast}{\partial a_i} \right)
\label{2.10}
\end{equation}
where $\frac{\partial}{\partial a_i}$ ($i=1,2,3$) denote the
generators of the transformation. In our case the translation
invariance gives rise to the two zero modes
\begin{equation}
\Psi_0^j = \left(\partial_j\varphi_c,\partial_j\varphi_c^\ast
\right), \qquad j=1,2,
\label{2.11}
\end{equation}
and the global $U(1)$ symmetry results in the third zero mode
\begin{equation}
\Psi_0^3 = \left(\frac{\partial}{\partial\theta}\varphi_c,
\frac{\partial}{\partial\theta}\varphi_c^\ast\right).
\label{2.12}
\end{equation}

Path integral quantization of static non--linear fields about
their classical configurations with BRST invariance provides
a systematic procedure to remove the degeneracy of the
action and therefore leads to a meaningful transition
amplitude. In order to see this we first of all elevate the parameters
$\vec{a}$ and $\alpha$ to new dynamical variables depending on time.
The field about the classical configuration can be written
\begin{equation}
\varphi(\vec{x},t)=\varphi_c'+\eta(\vec{x},t)
\label{2.13}
\end{equation}
where
\begin{equation}
\varphi_c'= e^{i\alpha(t)}\varphi_c(\vec{x}-\vec{a}(t))
\label{2.14}
\end{equation}
and $\eta(\vec{x},t)$ is considered as a small fluctuation about
$\varphi_c'$. The time dependence of $\alpha(t)$ and $\vec{a}(t)$
is determined by
the time--dependent Euler--Lagrange equations associated
with the Lagrangian (\ref{2.1}). In the following we shall consider
phase space functional integrals. The conjugate momenta of the
fluctuation fields $\eta$ and $\eta^\ast$ are defined as usual,
\begin{equation}
\pi = \frac{\partial{\cal L}}{\partial(\partial_0\eta)} =
\frac{\partial{\cal L}}{\partial(\partial_0\varphi)}, \quad
\pi^\ast = \frac{\partial{\cal L}}{\partial(\partial_0\eta^\ast)} =
\frac{\partial{\cal L}}{\partial(\partial_0\varphi^\ast)}
\label{2.15}
\end{equation}
where
\begin{eqnarray}
\frac{\partial{\cal L}}{\partial(\partial_0\varphi)} & = &
-2 \left( \partial_0\varphi \partial_i\varphi^\ast-\partial_i\varphi
\partial_0\varphi^\ast\right)\partial_i\varphi^\ast \label{2.16} \\
\frac{\partial{\cal L}}{\partial(\partial_0\varphi^\ast)} & = &
2 \left( \partial_0\varphi \partial_i\varphi^\ast-\partial_i\varphi
\partial_0\varphi^\ast\right)\partial_i\varphi \label{2.17}
\end{eqnarray}
The Hamiltonian density is
\begin{equation}
{\cal H} = \pi\partial_0\eta+\pi^\ast\partial_0\eta^\ast - {\cal L}.
\label{2.18}
\end{equation}
Regarding $\vec{a}$ and $\alpha$ as dynamical variables the definition
of conjugate momenta for $\vec{a}$ and $\alpha$ leads to constraints:
\begin{equation}
P_i = \frac{\partial L}{\partial \dot{a}_i} = \int d^2x
\left[ \pi\nabla_i\varphi_c e^{i\alpha}+c.c.\right], \qquad i=1,2,3.
\label{2.19}
\end{equation}
The constraints are:
\begin{equation}
\Phi_i=P_i-\int d^2x
\left[ \pi\nabla_i\varphi_c e^{i\alpha}+c.c.\right], \qquad i=1,2,3
\label{2.20}
\end{equation}
where $L=\int{\cal L}d^2x$ is the total Lagrangian and the compact
notations for $\dot{a}$ and $\nabla$ are defined by $\dot{a}=(\dot{a}_1,
\dot{a}_2,\dot{\alpha})$, $\nabla=(\partial_1,\partial_2,i)$, $\Phi=(\Phi_1,
\Phi_2,\Phi_3)$ and $P=(P_1,P_2,P_3)$. The total Hamiltonian is
correspondingly
\begin{equation}
H=P\cdot\dot{a}+\int d^2x \, {\cal H}.
\label{2.21}
\end{equation}

\section{Gauge transformations induced by constraints}

By direct computation we find the trivial first class constraints algebra
\begin{equation}
\{ \Phi_i,\Phi_j \} = 0
\label{3.1}
\end{equation}
where $\{,\}$ denotes a Poisson bracket.
The condition that the constraints be maintained in the course of
time, i.\ e.\
\begin{equation}
\{ \Phi_i, H \} = 0.
\label{3.2}
\end{equation}
can be verified by rewriting the Hamiltonian in the form
\begin{equation}
H=\Phi\cdot\dot{a}+H'
\label{3.3}
\end{equation}
where
\begin{equation}
H'=\frac{1}{2}\int d^2x\, (\pi\sqcap+\pi^\ast\sqcap^\ast)
- L_0
\label{3.4}
\end{equation}
and
\begin{equation}
\sqcap = \partial_0\varphi = \frac{\vec{A}^2\pi+\vec{A}\cdot\vec{A}^\ast\pi^\ast}
{2[(\vec{A}\cdot\vec{A}^\ast)^2-\vec{A}^{\ast 2}\vec{A}^2]}
\label{3.5}
\end{equation}
and
\begin{equation}
A_i=\partial_i\varphi, \qquad \vec{A}^2 = \sum_{i=1}^{2} \partial_i\varphi
\partial_i\varphi
\label{3.6}
\end{equation}
and $L_0 = \int {\cal L}_0 d^2x$ is the original static Lagrangian.
The $\dot{a}_i$ play the role of  Lagrange multipliers.

The constraints are all primary and first class as they should be. The
first class constraints generate the following gauge transformation with
time--dependent parameters $\Lambda_i(t)$:
\begin{eqnarray}
\delta\eta & = & \{\eta,\Lambda\cdot\Phi\} = -\Lambda\cdot\nabla
\varphi_c e^{i\alpha} \nonumber \\
\delta\eta^\ast & = & \{\eta^\ast,\Lambda\cdot\Phi\} = -\Lambda\cdot\nabla^\ast
\varphi_c^\ast e^{-i\alpha} \nonumber \\
\delta\pi & = & \delta\pi^\ast = \delta P = 0 \nonumber\\
\delta a & = & \{a, \Lambda\cdot\Phi\} = \Lambda
\label{3.7}
\end{eqnarray}
where we have used the notation $\Lambda\cdot\nabla=\Lambda_1\partial_1
+\Lambda_2\partial_2-i\Lambda_3$, $\Lambda\cdot\nabla^\ast
=\Lambda_1\partial_1+\Lambda_2\partial_2+i\Lambda_3$.
The first order Lagrangian
 \begin{equation}
L=P\cdot \dot{a} + \int(\pi\dot{\eta}+\pi^\ast\dot{\eta}^\ast)d^2x - H
\label{3.8}
\end{equation}
can be seen to be invariant under the gauge transformation (\ref{3.7})
induced by the constraints $\Phi$. We may add terms into the Lagrangian
(\ref{3.8}) consisting of
Lagrange multipliers $\lambda_i$
multiplied by the constraints $\Phi_i$ such that a new Lagrangian is
\begin{equation}
L^T=L-\lambda\cdot\Phi.
\label{3.9}
\end{equation}
Considering the Lagrange multipliers $\lambda_i$ as new dynamical variables
with conjugate momenta $P_{\lambda_i}$ which imply additional constraints
$p_{\lambda_i}=0$,
the action $S$ is invariant, i.\ e.\
\begin{equation}
\delta S = \delta\int L^T dt = 0
\label{3.10}
\end{equation}
under this new gauge transformation provided the additional constraints
are added to the generator, i.\ e.\ we have
\begin{equation}
\delta\lambda_i = \{\lambda_i, \Lambda\cdot\Phi +
\dot{\Lambda}\cdot p_\lambda\} = -\dot{\Lambda}_i.
\label{3.11}
\end{equation}

We have thus transformed the symmetries of the action with a
static classical field configuration into a gauge symmetry.
However, the corresponding total Hamiltonian $H^T$ is not gauge invariant
except on the subspace defined by the constraints. The next step is therefore
to enforce a BRST invariance after which the next
step is to fix the gauge symmetry while maintaining the invariance
in the sense of an enlarged BRST invariance. One should also note
that the Hamiltonian is not invariant under the transformation
(\ref{3.7}).

\section{BRST invariance and gauge fixing}

Since we have enlarged the phase space by elevating the Lagrange
multipliers $\lambda_i$ to dynamical variables we correspondingly
obtain three new momenta and constraints from $L^T$. Thus
\begin{equation}
P_\lambda = \frac{\partial L^T}{\partial\dot{\lambda}} = 0.
\label{4.1}
\end{equation}
The gauge transformation for the conjugate momenta of $\lambda$
is trivial
\begin{equation}
\delta P_\lambda=0.
\label{4.2}
\end{equation}
In order to break the gauge invariance of the action $S$ in (\ref{3.10})
and therefore to remove the degeneracy, we introduce the
anticommuting ghost variables $c=(c_1,c_2,c_3)$, $\bar{c}=
\bar{c}_1,\bar{c}_2,\bar{c}_3)$ and their respective canonical
conjugate momenta $\Pi_c=(\Pi_{c_1},\Pi_{c_2},\Pi_{c_3})$ and
$\Pi_{\bar{c}}=(\Pi_{\bar{c}_1},\Pi_{\bar{c}_2},\Pi_{\bar{c}_3})$,
such that	
\begin{equation}
\{c_i,\Pi_{c_j}\}_+=\delta_{ij}, \qquad
\{\bar{c}_i,\Pi_{\bar{c}_j}\}_+=\delta_{ij}
\label{4.3}
\end{equation}
with
\begin{equation}
\Pi_c\equiv L \frac{\stackrel{\leftarrow}{\partial}}{\partial\dot{c}}
= \dot{\bar{c}},
\qquad
\Pi_{\bar{c}}\equiv \frac{\vec{\partial}}{\partial\dot{\bar{c}}} L = \dot{c}
\label{4.3a}
\end{equation}
and vanishing Poisson brackets among $c$'s and $\Pi_c$'s.
The BRST charge $\Omega$ is
\begin{equation}
\Omega=c\cdot\Phi + P_\lambda\cdot\Pi_{\bar{c}}
\label{4.5}
\end{equation}
constructed such as to generate
the following BRST transformation corresponding to the gauge
transformation (\ref{3.7}):
\begin{eqnarray}
\delta\eta & = & \{\eta,\Omega\} = -c\cdot\nabla
\varphi_c e^{i\alpha} \nonumber \\
\delta\eta^\ast & = &  -c\cdot\nabla^\ast
\varphi_c^\ast e^{-i\alpha} \nonumber \\
\delta a & = & c \nonumber \\
\delta \bar{c} & = & P_\lambda  \nonumber \\
\delta \Pi_c & = & -\Phi
\label{4.4}
\end{eqnarray}
and other variations are zero.
The gauge symmetry of $L^T$ can be broken by adding to $L^T$ the
trivially BRST invariant term
\begin{eqnarray}
L_{\mbox{gf}} & = & \delta[\bar{c}\cdot\chi-\lambda\cdot\Pi_c]
\nonumber \\
& = & P_\lambda\cdot\chi - \bar{c}\cdot\{\chi,\Phi\}\cdot c
+ \Pi_{\bar{c}}\cdot\Pi_c+\lambda\cdot\Phi
\label{4.6}
\end{eqnarray}
where $\chi=(\chi_1,\chi_2,\chi_3)$ is the chosen gauge fixing
condition and is assumed to be independent of ghost variables and
constraints, e.\ g.\ $\{\chi,P_\lambda\}=0$, and
\begin{equation}
\bar{c}\cdot\{\chi,\Phi\}\cdot c \equiv
\sum_{i,j=1}^3 \bar{c}_i\{\chi_i,\Phi_j\} c_j.
\label{4.7}
\end{equation}
The effective Hamiltonian is
\begin{eqnarray}
H^{\mbox{eff}} & = & P\cdot \dot{a} + P_\lambda\cdot\dot{\lambda}
+ \Pi_c\cdot\dot{c} + \dot{\bar{c}}\cdot\Pi_{\bar{c}} +
\int(\pi\dot{\eta}+\pi^\ast\dot{\eta}^\ast)d^2x  - L^T - L_{\mbox{gf}}
\nonumber \\
& = &  P_\lambda\cdot\dot{\lambda} +  \bar{c}\cdot\{\chi,\Phi\}\cdot c
-\lambda\cdot\Phi+\Pi_c\cdot\Pi_{\bar{c}}+H.
\label{4.8}
\end{eqnarray}
We end up with the transition amplitude,
\begin{eqnarray}
\langle \varphi'(\vec{x}) |  \varphi(\vec{x}) \rangle & = &
\int_{\varphi(\vec{x})}^{\varphi'(\vec{x})} {\cal D}\{\varphi\}
{\cal D}\{\pi\}{\cal D}\{\varphi^\ast\}{\cal D}\{\pi^\ast\}{\cal D}\{P\}
{\cal D}\{a\}{\cal D}\{P_\lambda\}{\cal D}\{\lambda\} \nonumber \\
& & \times{\cal D}\{\Pi_c\}
{\cal D}\{c\}{\cal D}\{\Pi_{\bar{c}}\}{\cal D}\{\bar{c}\} e^{iS_B}
\label{4.9}
\end{eqnarray}
where the action is
\begin{eqnarray}
S_B & = & \int dt\,(L^T+L_{\mbox{gf}}) \nonumber \\
& = & \int_t^{t'} dt\,\left(
\Pi_c\cdot\Pi_{\bar{c}}-\bar{c}\cdot\{\chi,\Phi\}\cdot c
+\lambda\cdot\Phi+\chi\cdot P_\lambda + L \right).
\label{4.10}
\end{eqnarray}
Integrating out the ghost variables and $P_\lambda$, $\lambda$
is straightforward and yields
\begin{eqnarray}
\langle \varphi'(\vec{x}) |  \varphi(\vec{x}) \rangle & = &
\int_{\varphi(\vec{x})}^{\varphi'(\vec{x})} {\cal D}\{\varphi\}
{\cal D}\{\pi\}{\cal D}\{\varphi^\ast\}{\cal D}\{\pi^\ast\}{\cal D}\{P\}
{\cal D}\{a\} \nonumber \\
& & \times \left(\prod_k\delta[\Phi(k)]\delta[\chi(k)]\right)
\det\{\chi_i,\Phi_j\}
e^{i\int_t^{t'} Ldt}.
\label{4.11}
\end{eqnarray}
A natural choice for the gauge fixing functional $\chi$ is the following
in terms of the zero modes $\Psi_0^i$
\begin{equation}
\chi_i=\int[\eta\Psi_0^i+\eta^\ast\Psi_0^{\ast i}] d^2x
\label{4.12}
\end{equation}
since this corresponds to the socalled `unitary' gauge and requires the
fluctuation components along zero modes to vanish.

The matrix elements of the determinant $\det\{\chi_i,\Phi_j\}$ can be
evaluated by computing the Poisson brackets, i.\ e.\
\begin{equation}
\det\{\chi_i,\Phi_j\} = \int \left\{ [\eta\nabla_j (\Psi_0^i e^{i\alpha})
+ c.\ c.\ ] - [\Psi_0^i\Psi_0^j + c.\ c.\ ] \right\} d^2x
= -\delta_{ij} + {\cal O}(\eta)
\label{4.13}
\end{equation}
with appropriate normalization to $1$ of the zero modes.
The integration of the conjugate momenta $P_i$ of the collective
coordinates is readily carried out.

We then express the field $(\varphi,\varphi^\ast)$ as the classical
configuration plus quantum fluctuations $\eta$ as in (\ref{2.13}) and
then expand the transition amplitude in powers of $\eta$
and retain terms up to the
one--loop approximation. The final result of the transition amplitude
is
\begin{equation}
\langle \varphi'(\vec{x}) |  \varphi(\vec{x}) \rangle =
\int {\cal D}\{a\} e^{iS_c} I
\label{4.14}
\end{equation}
where $S_c$ is the classical action evaluated at the Skyrme--likesoliton
$\varphi_c$. The functional integral of the fluctuation is the factor
\begin{equation}
I=\int {\cal D}\{\eta\}{\cal D}\{\eta^\ast\}{\cal D}\{\pi\}{\cal D}\{\pi^\ast\}
e^{i\int dt\int\frac{1}{2}(\pi\sqcap+\pi^\ast\sqcap^\ast)d^2x}
\prod_i\delta[\chi(i)] \det(N) e^{i \int y^T \hat{M} y d^2x}
\label{4.15}
\end{equation}
where $y=(\eta,\eta^\ast)$ and the operator $\hat{M}$ is defined by
eq.\ (\ref{2.7}) and $N_{ij}$ is an element of the matrix $N$
defined by the normalization integral
\begin{equation}
N_{ij}=\int[\Psi_0^i\Psi_0^j + c.\ c.\ ]d^2x.
\end{equation}

Here, as previously \cite{9,10}, we observe that the coefficient of
$\bar{c}c$ in (\ref{4.10}), i.\ e.\ the mass of the ghosts, is the Poisson
bracket of constraints and gauge fixing conditions which in turn is the
Faddeev--Popov determinant in the Faddeev--Popov method.
The decoupling of ghosts in the leading loop approximation is
achieved only with the unitary gauge, which requires fluctuation components
in the directions of the zero modes to vanish.  

\section{Analysis of the transition amplitude in terms of eigenmodes of
the second variation operator}

One of the purposes of the BRST invariant gauge
fixing in the path--integral quantization of nonlinear fields about
classical configurations, as so far employed, is to remove the
degeneracy of the action which leads to ill defined functional integrals.
In order to demonstrate the latter explicitly we consider now $(1+1)$
dimensional $\phi^4$ field theory in order not to blur the main
points by complications of the Skyrme--like model.

In  $\phi^4$ field theory the canonical momentum integral of the
fluctuation field $\eta$ in eq.\ (\ref{4.15}) becomes a Gaussian
integral and therefore can be evaluated. The determinant is
\begin{equation}
\det N = \sqrt{M_0}
\label{4.16}
\end{equation}
where
\begin{equation}
M_0 = \int \left[ \frac{1}{2}\left(\frac{d\phi_c}{dx}\right)^2 + U(\phi_c)
\right] dx
\label{4.17}
\end{equation}
is the soliton mass, which with the use of the static equation of motion
is simply the normalization of the zero mode, i.e. $M_0 = \int (\frac{d\phi_c}{dx})^2dx$.

We let $\Psi_n$ be an eigenmode of the second variation operator
$\hat{M}$ such that
\begin{equation}
\hat{M}\Psi_n=E_n\Psi_n
\label{4.18}
\end{equation}
Expanding the fluctuation field $\eta$ in terms of the set $\{\Psi_n\}$,
we have
\begin{equation}
\eta = \sum_n c_n\Psi_n.
\label{4.19}
\end{equation}
We then change the variable of integration in the
path--integral from $\eta$ to $c_n$. The
path integral for the fluctuation then becomes
\begin{eqnarray}
I & = & \left|\frac{\partial \eta}{\partial c_n}\right| \int {\cal D}\{c_n\}
\delta[c_0] e^{i\int dt \sum c_n^2E_n} \nonumber \\
& = & \sqrt{M_0} \prod_{n\not= 0} \frac{1}{\sqrt{E_n}}
\label{4.20}
\end{eqnarray}
Since the transformation (\ref{4.19}) is linear the associated Jacobian
determinant is constant and can be factored out from the integration.
In our derivation all constant quantities have been dropped.
Here the BRST invariant gauge fixing plays only the role of bringing
in the factor $\delta[c_0]$ to remove the undefined integration over
the zero mode ${\cal D}\{c_0\}$ by replacing it by the integration
$da$.

It is interesting to observe that it suffices to consider only the static
Lagrangian without recalling the time dependence. We start from the
transition amplitude
\begin{equation}
\langle \phi'(\vec{x}) |  \phi(\vec{x}) \rangle =
\int {\cal D}\{\phi\} e^{iS_0}.
\label{4.21}
\end{equation}
After expansion about the classical configuration the classical
action can be factored out:
\begin{equation}
\langle \phi'(\vec{x}) |  \phi(\vec{x}) \rangle =
\int e^{iS_c} {\cal D}\{\eta\} e^{i\Delta S}
\label{4.22}
\end{equation}
and the remaining part of the action containing the selfadjoint
fluctuation operator $\hat{M}$ is
\begin{equation}
\Delta S = \int \eta\hat{M}\eta dx
\label{4.23}
\end{equation}
which is invariant under the field shift \cite{13}
\begin{equation}
\eta \rightarrow \eta + \Psi_0
\label{4.24}
\end{equation}
where $\Psi_0$ denotes the zero mode of operator $\hat{M}$. In order to
break this symmetry of the action we again add a BRST invariant
term to the Lagrangian. The BRST charge in the present case is defined
as
\begin{equation}
\Omega = -cP_a-c\pi\Psi_0(a)+\Pi_{\bar{c}}b
\label{4.25}
\end{equation}
where $P_a$ is the momentum conjugate to the translation parameter
$a$, and $\Psi_0$ is the associated zero mode, and $b$ is the Nakanishi--
Lautrup auxiliary field. The nonvanishing Poisson brackets of $a$,$P_a$
and the ghost variables are taken to be the canonical relations
\begin{eqnarray}
\{a,P_a\}_- & = & 1 \nonumber \\
\{\bar{c},\Pi_{\bar{c}}\}_+ & = & 1 \nonumber \\
\{c,\Pi_c\}_+ & = & 1.
\label{4.26}
\end{eqnarray}
The BRST charge generates the following transformations
\begin{eqnarray}
\delta \eta & = & \{\eta,\Omega\} = -c\Psi_0 \nonumber \\
\delta a & = & -c \nonumber \\
\delta \bar{c} & = & b \nonumber \\
\delta b & = & 0 \nonumber \\
\delta c & = & 0 \nonumber \\
\delta \Psi_0 & = & -c\frac{d\Psi_0}{dx}.
\label{4.27}
\end{eqnarray}
Next we add the following BRST invariant term to the fluctuation
Lagrangian
\begin{eqnarray}
L_B & = & \int \delta [\bar{c}\Psi_0\eta]dx - \frac{1}{2}b^2 \nonumber \\
& = & b\int\Psi_0\eta dx + \bar{c}c\int\Psi_0^2 dx +  \bar{c}c\int\eta
\frac{d\Psi_0}{dx}dx - \frac{1}{2}b^2
\label{4.28}
\end{eqnarray}
where the field $b$ can be replaced by the solution of its equation
of motion, i.\ e.\
\begin{equation}
b = \int\Psi_0\eta dx.
\label{4.29}
\end{equation}
Thus, $b$ is identified with the previous gauge fixing variable $\chi$.
The transition
amplitude with BRST invariant action is then
\begin{equation}
\langle \phi' |  \phi \rangle =
\int e^{iS_c} {\cal D}\{\eta\}{\cal D}\{a\}{\cal D}\{c\}
{\cal D}\{\bar{c}\}e^{i\int\eta\hat{M}\eta dx}
e^{i\left\{\frac{1}{2}\left[\int\Psi_0\eta dx\right]^2 +
\bar{c}c\int\left(\Psi_0^2-\Psi_0\frac{d\eta}{dx}\right)dx \right\} }.
\label{4.30}
\end{equation}
Integrating out ${\cal D}\{c\}$ and ${\cal D}\{\bar{c}\}$ we obtain
\begin{equation}
\langle \phi' |  \phi \rangle =
\int \sqrt{M_0} e^{iS_c} {\cal D}\{a\} \int {\cal D}\{\eta\}
e^{i\int\eta\hat{M}\eta dx} e^{i\frac{1}{2}\left[\int\Psi_0\eta dx\right]^2}.
\label{4.31}
\end{equation}
Expanding $\eta$ in terms of eigenmodes of the operator $\hat{M}$
and a changing integration variables from ${\cal D}\{\eta\}$ to
${\cal D}\{c_n\}$, we find that integration over the zero mode
coefficient, ${\cal D}\{c_0\}$, becomes a Gaussian integral and can be
carried out. The final result of the transition amplitude is seen to be
the same as (\ref{4.20}) of the previous method.
We therefore conclude that the two treatments
are equivalent in the one loop approximation.
Of course, since the invariance under the shift (59) occurs only in
the one--loop approximation, this is also the limit of its validity.  
Nonetheless this is an interesting observation which again
indicates a novel property of the zero mode.

This second method is
particularly useful in the explicit calculation of quantum mechanical tunneling
effects with the instanton method. Research along this direction is in progress.

\section{Global symmetry and Ward--Takahashi (WT) identity}

In the previous section we evaluated in the one--loop approximation
the transition amplitude of the complex scalar field in the background
of Skyrme--like solitons without external sources. The symmetries of
the action were broken to obtain well defined functional integrals.
The global $U(1)$ symmetry is a new feature of the model. We therefore
add some comments and derive the WT identities related to this
symmetry \cite{24}. To this end we start from the generating functional
of correlation functions:
\begin{equation}
Z(J) = \int {\cal D}\{\varphi\}{\cal D}\{\varphi^\ast\} \exp \left[ iS_0
(\varphi,\varphi^\ast) + i\int d^2x (J\varphi + J^\ast\varphi^\ast)\right]
\label{6.1}
\end{equation}
where $S_0$ is the action defined by the static Lagrangian (\ref{2.3})
and $J$ is the current of external sources. The classical action $S_0$ is
invariant under global $U(1)$ transformations $\varphi'=e^{i\alpha}\varphi$,
$\delta\varphi=i\alpha\varphi$, and therefore
\begin{equation}
\int d^2x \left(i\alpha\varphi\frac{\delta S_0}{\delta\varphi}-
i\alpha\varphi^\ast\frac{\delta S_0}{\delta\varphi^\ast}\right) = 0
\label{6.2}
\end{equation}
The measures of integration ${\cal D}\{\varphi\}$ and
${\cal D}\{\varphi^\ast\}$ in the functional integral (\ref{6.1}) are
invariant under the unimodular transformation $\varphi'=e^{i\alpha}\varphi$.
Demanding $\delta Z(J) = 0$ leads to the equation
\begin{equation}
i\alpha\int {\cal D}\{\varphi\}{\cal D}\{\varphi^\ast\}
\int d^2x (J\varphi - J^\ast\varphi^\ast)\exp\left[ iS_0 +
i\int d^2x (J\varphi + J^\ast\varphi^\ast)\right] = 0
\label{6.3}
\end{equation}
We then have
\begin{equation}
\int d^2x \left( J\frac{\delta Z(J)}{\delta J} -
J^\ast\frac{\delta Z(J)}{\delta J^\ast} \right)= 0
\label{6.4}
\end{equation}
or
\begin{equation}
\int d^2x \left( J\frac{\delta W}{\delta J} -
J^\ast\frac{\delta W}{\delta J^\ast} \right) = 0
\label{6.5}
\end{equation}
and
\begin{equation}
\int d^2x \left( J\frac{\delta \Gamma}{\delta J} -
J^\ast\frac{\delta \Gamma}{\delta J^\ast} \right) = 0
\label{6.6}
\end{equation}
where $W$ is the generating functional of the connencted correlation
function while $\Gamma$ denotes proper vertices.

\subsection*{Acknowledgement}

This work was supported in part by the European Union under the
Human Capital and Mobility programme. J.--G.~Zhou also
acknowledges support of the A.\ v.\ Humboldt Foundation,
J.--Q.Liang support of the Deutsche Forschungsgemeinschaft and D.\ H.\
Tchrakian partial support by the Irish and German Science
cooperation coordinators EOLAS (Ireland) and GKSS (Germany).

\end{document}